\documentclass[aps,twocolumn,superscriptaddress,psfig,showpacs,prl]{revtex4}
\usepackage{epstopdf} 
\usepackage{epsfig}

\begin{document}

\title{Viscoelastic properties of attractive and repulsive colloidal glasses}
\author{Antonio M. Puertas}
\email[Corresponding author: ]{apuertas@ual.es}
\affiliation{Group of Complex Fluids Physics, Department of Applied Physics, University of Almeria, 04120 Almeria, Spain}
\author{Emanuela Zaccarelli}
\affiliation{Dipartimento di Fisica and INFM Udr and SOFT: Complex Dynamics in Strucured Systems, Universita` di Roma La Sapienza, Piazzale Aldo Moro 5, I-00185 Rome, Italy}
\author{Francesco Sciortino}
\affiliation{Dipartimento di Fisica and INFM Udr and SOFT: Complex Dynamics in Strucured Systems, Universita` di Roma La Sapienza, Piazzale Aldo Moro 5, I-00185 Rome, Italy}
\date{\today}

\begin{abstract}
We report a numerical study of the shear viscosity and the frequency
dependent elastic moduli close to dynamical arrest for a model of
short-range attractive colloids, both for the repulsive and the
attractive glass transition. Calculating the stress autocorrelation
functions, we find that density fluctuations of wavevectors close to
the first peak in the structure factor control the viscosity rise on
approaching the repulsive glass, while fluctuations of larger
wavevectors control the viscosity close to the attractive glass.  On
approaching the glass transition, the viscosity diverges with a power
law with the same exponent as the density autocorrelation time.
\end{abstract}


\maketitle

Colloidal dispersions, in addition to their technological relevance,
play an important role in the development of basic physical
sciences, in particular in the fascinating field of formation of
disordered arrested states, glasses and gels. The possibility of
tailoring shape, size and structure of the colloidal particles makes
it possible to design specific colloidal interaction potentials and to
use light scattering to directly probe even single particle
motion. Recently, the application of
experimental\cite{pham02,barsh02,chen03} and theoretical
tools\cite{bergenholtz99,fabbian99,dawson00} to the analysis of the
glass transition in short-ranged attractive colloids, has shown an
extremely rich scenario, with no analogue in atomic
systems\cite{sciortino02}. The standard packing-driven hard-sphere
glass transition transforms -- discontinuously in some cases -- into a
novel-type of glass transition driven by the short-range
attractions. The connection between this attractive transition and
gelation is still matter of debate.

Previous numerical work on short-range attractive colloids has mostly
focused on the behavior of single particle diffusion and of the time
dependence of density autocorrelation function $\Phi_q(t)$ close to
dynamical arrest. Despite the strong link with experiments and the
relevance to industrial applications, the numerical evaluation of the
viscosity, $\eta$, and viscoelastic properties $\tilde \eta(\omega)$
has not been performed, since significant computational effort is
required for accurate calculation of $\tilde \eta(\omega)$, even more
for states close to dynamical arrest. The behavior of $\eta$ close to
dynamical arrest in colloidal systems is under theoretical and
experimental
investigation\cite{fuchs,harward,cheng02,mallamace04}. Measurements in
dense hard sphere colloids show a divergence of $\eta$ in the vicinity
of the repulsive glass point, its exact location and functional
form being still under debate \cite{cheng02,fuchs_faraday03}. For
colloidal gels, a power law divergence has been reported in connection
to the gel transition \cite{shah03}.  The Mode Coupling Theory for
glass transitions\cite{mct} (MCT), the theoretical framework which
anticipated the presence of novel dynamic phenomena in short-range
attractive colloids\cite{bergenholtz99,fabbian99,dawson00} predicts
asymptotic power law divergence of $\eta$ with the distance to the
transition, with identical exponents to the divergence of the time
scale, $\tau$, and the inverse of the self diffusion coefficient
$1/D_0$.

In this Letter we report an extensive numerical study of the
viscoelastic behavior of a colloidal system, approaching both the
repulsive and the attractive glass transition.  We calculate the
stress autocorrelation function, as well as the viscosity and the
elastic modulii. We find that, close to both transitions, $\eta$
diverges with a power law, with the same exponent as the density
relaxation time, but different from that of $1/D_0$. Moreover, we
provide evidence that the rise of $\eta$ is controlled by density
fluctuations of wavevectors around the first peak in the structure
factor close to the repulsive glass, and by fluctuations of larger
wavevectors close to the attractive glass.

Our system is composed of $1000$ particles interacting via a steep
repulsive ($r^{-36}$) potential complemented by the Asakura-Oosawa
(AO) short-range attractive potential \cite{likos}.  The attraction
strength is measured in units of the polymer volume fraction, $\phi_p$
and density is reported as colloid volume fraction, $\phi_c \equiv 4/3
\pi a [a^2+\delta^2] \rho$, with $\rho$ the number density, and the
range of the interaction is $\xi=0.1a$ \cite{potential}.  Dynamics in
colloidal systems can be conveniently described by Brownian dynamics
(BD), which accounts for the presence of solvent (not treating the
full hydrodynamic interactions). Unfortunately, BD simulations are
effectively less efficient than Newtonian dynamics (ND) due to the
slower microscopic time scale. Since as far as slow dynamics is
concerned, BD and ND are equivalent~\cite{gleim98,voigtmann04} we
mostly work with the more efficient ND but perform additional
simulations with BD at high friction coefficient, $\gamma=50$
\cite{bd}, to confirm the independence on the short-time dynamics.
\begin{figure}[h]
\psfig{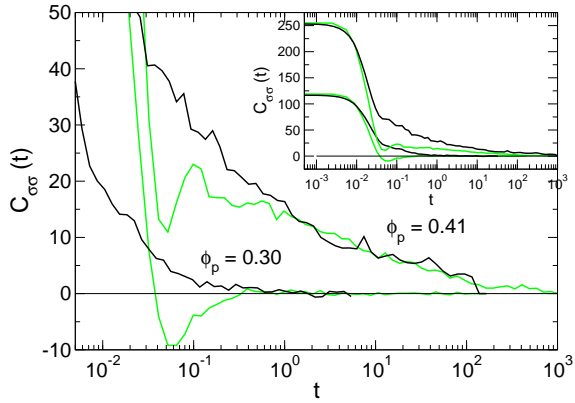}
\caption {\label{nd-bd} Stress correlation functions for ND (grey
lines) and BD (black lines), time rescaled, for $\phi_p=0.30$ and
$0.41$ as labeled. For both states, $C_{\sigma\sigma}^{BD}(t)$ has
been rescaled onto $C_{\sigma\sigma}^{ND}(t)$ by a factor 6.25. Inset:
Full scale stress correlation functions for ND and BD (without time
rescaling).}
\end{figure}
Dynamics in the chosen model system have been studied
previously\cite{puertas03,voigtmann04}. In the absence of polymers
($\phi_p=0$) and with no repulsive barrier, on increasing the particle
packing fraction a repulsive glass transition is observed at $\phi_c^G
\simeq 0.594$. At fixed $\phi_c=0.40$, on increasing $\phi_p$, an
attractive glass transition is observed at $\phi_p^G\simeq
0.4265$. MCT analysis for $\Phi_q(t)$ and $D_0$ has been performed
previously for both paths \cite{puertas03, voigtmann04}.

\noindent
The shear viscosity $\eta$ is given by the Green-Kubo relation
\begin{equation} 
\eta\: \equiv \:\int_0^{\infty}
dt\,C_{\sigma\sigma}(t)\:=\:\frac{\beta}{3V}\int_0^{\infty}
dt\,\sum_{\alpha < \beta} \langle \sigma^{\alpha\beta}(t)
\sigma^{\alpha\beta}(0) \rangle,
\label{eq:eta}
\end{equation}
\noindent which expresses $\eta$ as the integral of the correlation
function of the non-diagonal terms of the microscopic stress tensor,
$\sigma^{\alpha\beta}\:=\:\sum_{i=1}^Nm
v_{i\alpha}v_{i\beta}\,-\,\sum_{i<j}^N \frac{r_{ij\alpha}
r_{ij\beta}}{r_{ij}} V'(r_{ij})$, where $V$ is the volume of the
simulation box, $v_{i\alpha}$ is the $\alpha$-th component of the
velocity of particle $i$, and $V'$ is the derivative of the total
potential.

The inset in Fig. \ref{nd-bd} shows $C_{\sigma\sigma}(t)$ for two
state points with $\phi_c=0.40$: $\phi_p=0.30$ and $\phi_p=0.41$, both
for ND and BD. In the case of ND, clear oscillations in
$C_{\sigma\sigma}(t)$ at small time ($t<0.5$) are observed, 
caused by motion in the attractive well. 
These oscillations are completely damped in
the BD case, confirming that this part of the decay of the correlation
function has a microscopic (dynamic-dependent) origin and that short
time oscillations are not to be expected in real colloidal
systems. 
When $\phi_p$ is increased, an additional slow relaxation process is
present. Interestingly, the BD $C_{\sigma\sigma}(t)$ can be
time-rescaled to collapse at long $t$ on ND data (main figure),
confirming that the long time dynamics of the system does not depend
on the microscopic dynamics, in agreement with previous studies in
glass forming atomic systems\cite{gleim98}. Therefore, we use ND
hereafter to analyse the long time decay of $C_{\sigma \sigma}(t)$ and
the viscosity close to both transitions.

\begin{figure}[ht]
\psfig{file=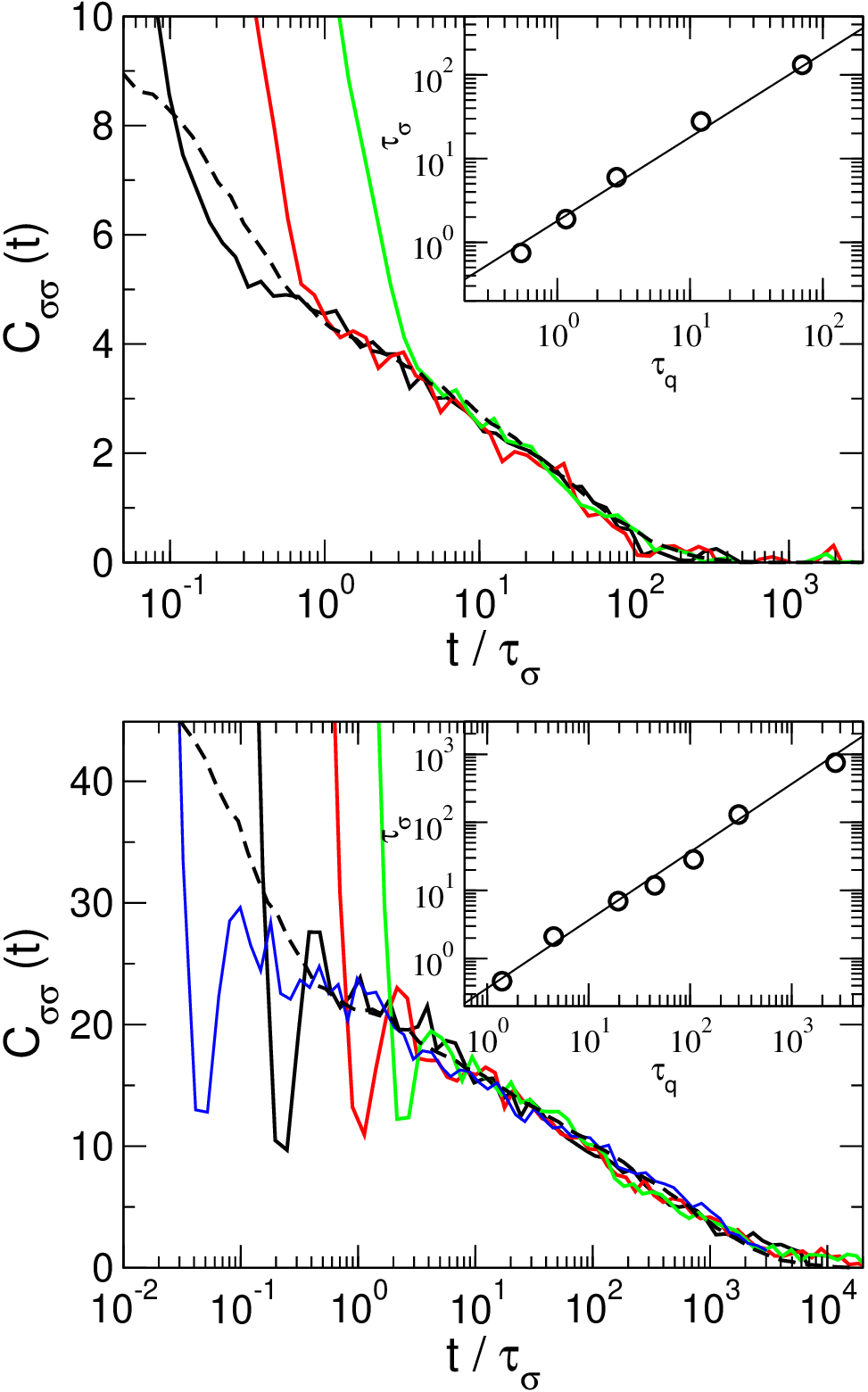,width=7.5cm}
\caption{\label{figure1} Stress correlation functions for different
states approaching the glass transitions. Upper panel: Repulsive glass
transition. From right to left: $\phi_c=0.55, 0.57, 0.58$. Lower
panel: Attractive glass transition. From right to left:
$\phi_p=0.405,0.41,0.415,0.42$ and $\phi_c=0.40$. Dashed lines are
$\Phi_q(t)^2$ for $qa=4$ at $\phi_c=0.58$ (repulsive glass) and
$qa=11$ at $\phi_p=0.415$ (attractive glass). Insets: Time scale from
the stress correlator vs. time scale from density correlator at the
wavevectors given above; $C_{\sigma\sigma}(\tau_{\sigma})=1$
(repulsive glass), $C_{\sigma\sigma}(\tau_{\sigma})=5$ (attractive
glass) and $\Phi_q(\tau_q)=f_q/e$.}
\end{figure}

Fig.~\ref{figure1} shows the stress correlation function,
$C_{\sigma\sigma}(t)$, for different $\phi_c$ or $\phi_p$ values on
approaching the repulsive (upper panel) and attractive (lower panel)
glass transitions. Correlation functions have been superimposed by a
suitable scaling of time, $\tau_{\sigma}$ (insets of
Fig. \ref{figure1}). On approaching the transition,
$C_{\sigma\sigma}(t)$ slows down by several orders of magnitude, with
no appreciable change in its shape at long-time. A stretched
exponential can be fitted to the decay with an exponent $\beta=0.48$
and $\beta=0.25$ for the repulsive and attractive glasses,
respectively. As can be seen from Eq.\ref{eq:eta}, the slowing down of
$C_{\sigma\sigma}(t)$ is responsible of the rise of $\eta$. 

The scaling property of $C_{\sigma\sigma}(t)$ at long times is
reminiscent of the scaling properties of $\Phi_q(t)$. Indeed, within
MCT, $C_{\sigma\sigma}(t)$ can be related to $\Phi_q(t)$, by means of
\cite{bengtzelius84,nagele}:
\begin{equation}
C_{\sigma\sigma}(t)\:=\:\frac{k_BT}{60 \pi^2} \int_0^{\infty} dq\,q^4
\left[ \frac{d\,\ln S_q}{dq}\,\Phi_q(t) \right]^2
\label{csigmasigmaMCT}
\end{equation}
\noindent where $S_q$ is the static structure factor. Accordingly, the
$\phi_c$ or $\phi_p$ dependence of $\tau_{\sigma}$ should be identical
to the $\phi_c$ or $\phi_p$ dependence of $\tau_q$, the characteristic
time of $\Phi_q$: $\Phi_q(\tau_q)=f_q/e$, with $f_q$ the
non-ergodicity parameter. The two insets of Fig.~\ref{figure1} show a
plot of $\tau_{\sigma}$ vs. $\tau_q$, parametric in $\phi_c$ or
$\phi_p$. For both types of glass transitions, the time scales
obtained by rescaling $C_{\sigma\sigma}(t)$ are proportional to that
found rescaling $\Phi_q(t)$ (continuous lines).

The dominant contribution to the $q$-integral in
Eq.~\ref{csigmasigmaMCT}, for times in the $\alpha$-decay of
$\Phi_q(t)$, can be revealed studying the function $q^4 \left[ d\,\ln
S_q/dq\,f_q\right]^2$. This function oscillates with $S_q$ (out of
phase $\pi/2$), but its envelope shows a maximum close to the nearest
neighbour peak of $S_q$ for the repulsive glass transition ($qa
\approx 3.75$), and at higher $q$ for the attractive one (the exact
value depending on the details of the interaction potential). For the
AO potential, using Percus-Yevick closure, the dominant wavevector 
is $qa\approx 13$, although the
distribution is very wide. Interestingly, the wavevectors driving both
the repulsive and the attractive ideal MCT glass transitions, as
observed from the density correlators, are also the dominant ones in
the calculation of $C_{\sigma\sigma}(t)$.

Guided by Eq.\ref{csigmasigmaMCT}, we evaluate $\Phi_q$ from the
simulations and compare $\Phi_q^2(t)$ with $C_{\sigma\sigma}(t)$ for
different wavevectors, allowing for a scaling factor only in the
amplitude. The best agreement is obtained for $qa=4$ in the repulsive
glass transition and $qa=11$ in the attractive glass (dashed lines in
Fig.~\ref{figure1}), in good agreement with MCT predictions. The
comparison shows that the long time decay of the stress correlation
function, within the accuracy of our numerical results, is adequately
described by the dominant wavevector in Eq.~\ref{csigmasigmaMCT}
\cite{nota}.

\begin{figure}[h]
\psfig{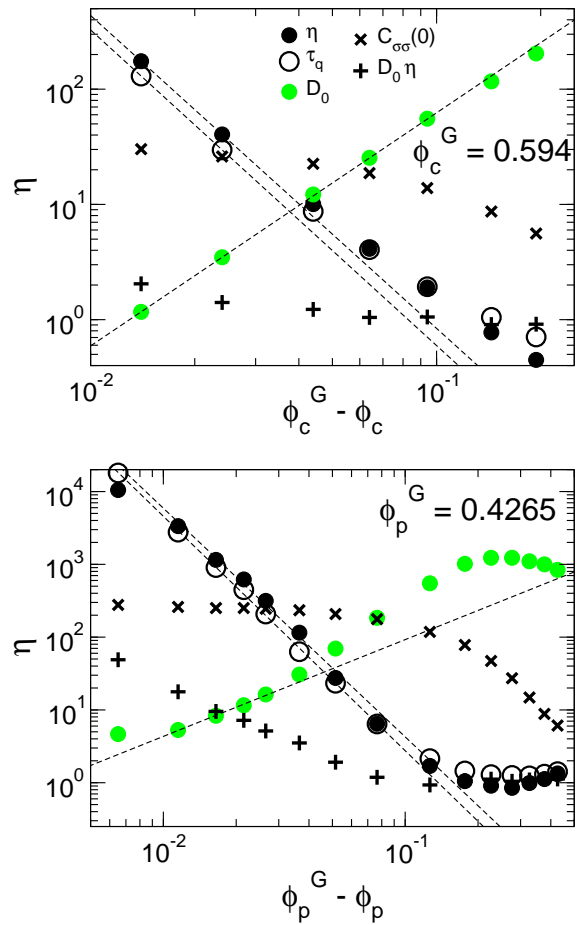}
\caption {\label{figure2} Viscosity, $\eta$, time scale at $q=3.5$,
diffusion coefficient ($D_0\times\:10^3$) and $C_{\sigma\sigma}(0)$
for the repulsive glass transition (upper panel) and the attractive
one (lower panel), as labeled. Note that $\eta D_0$ diverges in the
transition point. The lines are the power law fittings to $\eta$,
$\tau_q$ and $D_0$ (see Table I).}
\end{figure}

We now turn to the study of $\eta$, which is more efficiently
calculated numerically using the Einstein relation \cite{allen87}:
\begin{equation}
\eta\:=\:\frac{\beta}{6V} \lim_{t\rightarrow \infty} \frac{1}{t}
\langle \Delta A(t)^2 \rangle,
\end{equation}
\noindent where $\Delta A(t)$ is the integral from $0$ to $t$ of the
three off-diagonal terms of the stress tensor.  Fig.~\ref{figure2}
shows the behavior of $\eta$ approaching the repulsive and the
attractive glass transitions, as a function of the distance to the
transition points, $\phi_c^G$ and
$\phi_p^G$\cite{puertas03,voigtmann04}. For both the repulsive and
attractive glass transitions, $\eta$ is well described by a power law.
In the attractive case, $\eta$ shows a minimum for intermediate
$\phi_p$, arising from the competition of the two competing arrest
mechanisms\cite{pham02,foffi02}.

Fig.~\ref{figure2} also shows $\tau_q$ for $qa=3.5$ (the nearest
neighbour peak in $S_q$), $D_0$, $C_{\sigma\sigma}(0)$ and $\eta
D_0$. For both transitions, both $\tau_q$ and $D_0$ are linear in
log-log and hence can also be well described by power laws. While
$\eta \sim \tau_q$, as confirmed by the constant ratio $\eta/\tau_q$
(not shown), the product $\eta D_0$ shows a clear trend, indicating
that the power-law exponents are very similar for $\eta$ and $\tau_q$
but different for $D_0$. The best-fit values are reported in Table I
for both transitions.
\begin{table}[h]
\begin{ruledtabular}
\begin{tabular}{lccc}
 & $\gamma_{\eta}$ & $\gamma_{\tau}$ & $\gamma_{D_0}$ \\ \hline
Repulsive glass &  2.72 & 2.74 & 2.02 \\
Attractive glass & 3.14* & 3.23 & 1.33 \\ 
\end{tabular}
\end{ruledtabular}
\caption{Exponents obtained in fitting $\eta$, $\tau$ and $1/D_0$ to
$(\phi_c^G-\phi_c)^{-\gamma}$, for the repulsive glass and
$(\phi_p^G-\phi_p)^{-\gamma}$ for the attractive one. (*) The last
point, $\phi_p=0.42$, is not considered in the fitting, due to the
large error in determining $\eta$.}
\end{table}
\begin{figure}[h]
\psfig{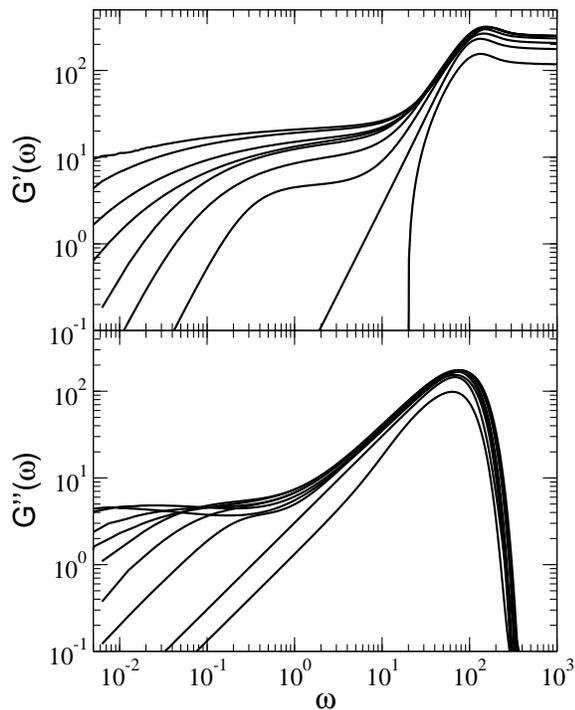}
\caption {\label{moduli} Elastic (upper panel) and viscous (lower
panel) modulus for different states approaching the attractive glass
transition: From right to left
$\phi_p=0.30,0.35,0.375,0.39,0.40,0.405,0.41,0.415,0.42$.}
\end{figure}
The difference between the exponents $\gamma_{\eta}$ and
$\gamma_{D_0}$ indicates that the Stokes-Einstein relationship breaks
down not only in the form $D_0-\tau$, as already shown in many
different systems, but also in the form $D_0-\eta$. This breakdown,
much more evident in the attractive glass case, is not consistent with
MCT, although the theory correctly predicts the similarity between the
exponents of $\tau_q$ and $\eta$. The power-law divergence of $\eta$
at the attractive glass transition agrees with recent experimental
observations in colloidal gels \cite{shah03}. It is also consistent
with recent interpretation of data for the repulsive glass transition
\cite{fuchs_faraday03}, although this analysis is not conclusive,
since an exponential divergence around $\phi_m=0.64$ has also been
suggested \cite{cheng02}. The constant $C_{\sigma\sigma}(0)$ is the
elastic modulus at infinite frequency $G_{\infty}'$. As both
transitions are approached, it tends to a finite value without showing
any divergence, also in agreement with MCT.

To make direct contact with experiments, we calculate $\tilde
G(\omega)\equiv i\omega \tilde \eta(\omega)=G'(\omega)+iG''(\omega)$,
where $\tilde \eta(\omega)$ is the Fourier transform of
$C_{\sigma\sigma}(t)$.
To minimize numeric noise, we fit $C_{\sigma\sigma}(t)$ using
functional forms that capture the fast short-time behaviour\cite{note}
and a long-time stretched exponential decay.  Fig.~\ref{moduli} shows
the resulting $\tilde G(\omega)$ for the attractive glass transition
case. Similar qualitative features are observed for the repulsive
case. An incipient plateau at low frequencies in $G'(\omega)$ and a
secondary maximum in $G"(\omega)$, leaving a minimum at intermediate
frequencies, appear on approaching $\phi_p^G$. Both features are
consistent with MCT predictions \cite{fuchs99,dawson01} and observed
in experiments \cite{mason95,shah03,mallamace04}. The short-time
behavior of $C_{\sigma\sigma}(t)$ causes the sharp increase in
$G'(\omega)$ at $\omega \sim 10^1$.

In summary we have shown that in short range attractive colloids
$\eta$ diverges following a power law as the transition points are
approached, with the same exponent as the time scale $\tau_q$, but
different from that of $D_0$. On approaching $\phi^G$, the
slow-decaying part of $C_{\sigma\sigma}(t)$ can be time rescaled onto
a master function which has the same shape as $\Phi_q(t)^2$ at a
specific $q$.  For the case of repulsive glass, the inverse $q$ value
corresponds to the nearest neighbor distance, while, in the case of
attractive glass, it corresponds to distances comparable to the (short)
range of attraction. Finally, the frequency dependence the elastic
moduli is consistent, for both transitions, with predictions based on
MCT and with recent experiments on colloidal gels and glasses.

We thank P. Tartaglia 
for useful discussions and
M. Fuchs, M. Cates and W. G\"otze for careful reading of the
manuscript. A.M.P. thanks the Dipartimento di Fisica at the
Universit\`a di Roma "La Sapienza", where this work was carried out,
for hospitality and the Spanish Ministerio de Educacion, Cultura y
Deporte for financial support. E.Z. and F.S. acknowledge support
from Miur FIRB, Cofin and MRTN-CT-2003-504712.


\begin{thebibliography}{99}
\bibitem{potential} The system is polydisperse to avoid
crystallization (flat distribution, average radius $a=1$, and
half-width $\delta=0.1 a$), and a long-range repulsive barrier is
added to inhibit liquid-gas separation at high attraction
strength. The specific shape of the potential can be found in
\protect\cite{puertas03}.
\bibitem{bd} Such high friction coefficient is needed to decouple
momemtum relaxation from the effect of the attractions.
\bibitem{nota} The identification of the decay of $C_{\sigma\sigma}$
with a single $\Phi_q$ is of course an approximation. Indeed,
different $q$ are characterized by stretched exponential decay with
different values of $\tau_q$ and $\beta_q$.  Slow decaying
contributions ($q$ around the nearest neighbor peak of $S(q)$) will
always control the decay of $C_{\sigma\sigma}$ at very long times for
both glass transitions. In the present simulations, however, such
effect is not observed within the accuracy of our data.
\end{thebibliography}
\end{document}